\documentclass[11pt]{article}

\pdfoutput=1

\usepackage{epsfig,natbib,amsmath,float,color,fullpage,amssymb,rotating,bbm}
\usepackage{blindtext}
\usepackage{graphicx}
\usepackage{algorithm2e}
\usepackage{array}
\usepackage{layout}
\usepackage{multicol}

\newcommand{\quotes}[1]{``#1"}
\def\N{\mathcal{N}}
\def\IG{\mathcal{IG}}
\def\Ga{\mathcal{G}}

\def\NIG{\mathcal{NIG}}
\def\M{\mathcal{M}}
\def\given{|} 

\begin{document}  
	\title{Sequential Monte Carlo Smoothing with Parameter Estimation}

	\author{Biao Yang\thanks{Ph.D. Candidate, Department of Statistics, George Washington University (yangbiao@gwu.edu).},
	           Jonathan R. Stroud\thanks{Associate Professor, McDonough School of Business, Georgetown University (jrs390@georgetown.edu).}, 
		  and Gabriel Huerta\thanks{Professor, Department of Mathematics and Statistics, University of New Mexico (ghuerta@stat.unm.edu). G.~Huerta was partially funded by the U. S. Department of Energy Office of Science, Biological and Environmental Research Regional and Global Climate Modeling under Award Number DE-SC0010843.}}

	\maketitle
	
\begin{abstract}
	In this paper, we propose two new Bayesian smoothing methods for general state-space models with unknown parameters. The first approach is based on the particle learning and smoothing algorithm, but with an adjustment in the backward resampling weights. The second is a new method combining sequential parameter learning and smoothing algorithms for general state-space models. This method is straightforward but effective, and we find it is the best existing Sequential Monte Carlo algorithm to solve the joint Bayesian smoothing problem.   We first illustrate the methods on three benchmark models using simulated data, and then apply them to a stochastic volatility model for daily S\&P 500 index returns during the financial crisis.
\end{abstract}

\vfill

\section{Introduction}
   
   State-space models are a powerful tool for handling nonlinear, non-Gaussian time series. This general class of models is widely used in many fields, including finance, ecology, biology and engineering.  Over the last few decades, Sequential Monte Carlo (SMC) methods have become extremely popular for sequential state and parameter estimation in state-space models.  These methods, however, have been largely ignored for Bayesian smoothing (i.e., retrospective analysis).  Smoothing presents computational challenges because the target posterior distribution is often high-dimensional and intractable.   In this paper, we propose two new SMC algorithms that overcome these challenges.

Markov chain Monte Carlo (MCMC) methods are the most common approach to Bayesian smoothing.   \cite*{Carlin:1992} introduced the first MCMC approach for nonnormal and nonlinear models. \cite*{Carter:1994} and \cite*{Fruhwirth:1994} proposed the forward-filtering, backward-sampling (FFBS) algorithm and \cite*{Piet:1995} introduced the related simulation smoother for conditionally Gaussian models. The FFBS is an efficient block sampler that draws the states jointly given the parameters for linear, Gaussian state-space models. \cite{Shephard:1997} and \cite{Gamerman:1998} provided block sampling algorithms for non-Gaussian and exponential family measurement models, respectively.  \cite*{Geweke:2001} proposed Metropolis-within-Gibbs algorithms for nonlinear and non-Gaussian state-space models, and \cite*{Stroud:2003} proposed a block sampling algorithm for nonlinear models with state-dependent variances. \cite*{Niemi2010} solved the nonlinear and non-normal case by sequential approximation of filtering and smoothing densities using normal mixtures.  
    
   Particle filtering is a sequential Monte Carlo method that has also been widely used for state estimation in state-space models and has been successful in many simulation studies and real data problems. The idea was first introduced by \cite*{Gordon:1993} with the name \quotes{bootstrap filter.}  Then,  \cite*{Pitt:1999} improved this by introducing the auxiliary particle filter. However, the problem of dealing with unknown parameters in Sequential Monte Carlo methods is not fully resolved. \cite*{Kitagawa:1998} proposed including the parameters into the state vector and proposed a particle filter on the augmented state vector.  On the other hand, \cite*{West:2001} use a kernel smoothing density for the static parameters to avoid over-dispersion problems. This filter algorithm remains to be the most general method for sequential state and parameter estimation. Both \cite*{Storvik:2002} and \cite*{Fearnhead:2002} discussed generating samples of the parameters from the filtering distribution in situations where sufficient statistics for $\theta$ are available.  In this case, the samples of parameters simulated at time $t$ do not depend on values simulated at previous times and the problem of impoverishment is mitigated.  A comprehensive review of parameter estimation for state-space models was recently given by \cite*{kantas:2015}, in which both maximum likelihood methods and Bayesian methods were discussed.    
   
   In addition to the filtering problem, in which state estimation is conditional on the data available at time $t$, Sequential Monte Carlo methods can also be applied to state smoothing. In smoothing problems, we estimate the states conditional on all the observations. \cite*{Kitagawa:1996} introduced the idea of smoothing by storing the state vector, in which the smoothing process is realized by resampling the filtered particles within the smoothing window. But as time evolves and the smoothing window width increases in size, the smoothing samples at the start of the time series will degenerate to a single path. Other smoothing algorithms include: the forward-backward smoother of \cite*{Godsill:2004}, in which a backward recursion is included and the forward filter particles are reweighted; the two-filter smoother of \cite*{Kitagawa:1996}; the generalized two-filter smoother of \cite*{Doucet:2010}; and the new $O(N)$ and $O(N^2)$ smoothing algorithms of \cite*{Fearnhead:2010}.

   All of the sequential Monte Carlo smoothing algorithms discussed above are based on the assumption that the fixed parameters are known. Research on particle smoothing with unknown parameters is limited. The particle learning and smoothing (PLS) algorithm of \cite*{Carvalho:2010} is one of the most well-known methods in this area. In their smoothing algorithm, however, the dependency between states and the parameters is ignored, which results in a failure of their smoothing algorithm at the beginning of the time series. In this paper, we take the dependency of state and parameters into consideration and adjust the resampling weights in the backward pass. In addition, we propose a new smoothing algorithm, in which we apply a forward-backward smoother on each parameter drawn from the last filter step and get a corresponding smoother sample.   This provides smoothed samples of the states while accounting for parameter uncertainty.

  There are three main advantages of our approach relative to existing methods.  First, our refiltering algorithm is the only Sequential Monte Carlo method to provide an ``exact" solution to the Bayesian smoothing problem as the number of particles goes to infinity.   Second, our smoothing algorithms can be easily parallelized, since communication between processors is minimal. Third, unlike MCMC approaches, marginal likelihood and Bayes' factors can be accurately computed at each time \citep{Carvalho:2010}, which is useful for sequential model comparison and model selection. 
  
In addition, we find empirical evidence that the posterior dependence between states and parameters decreases as time $t$ goes to infinity.  This suggests the possibility of new algorithms for on-line Bayesian state and parameter estimation that exploit this independence.  

\newpage
   This article is organized as follows. In Section 2, we give a brief review for particle filtering and smoothing algorithms. Two new smoothing algorithms are proposed in Section 3. In Section 4, the two new smoothing algorithms and PLS are tested on three models: an AR(1) plus noise model with three unknown parameters, a nonlinear growth model with five unknown parameters and a chaotic model with three unknown parameters. Finally, in Section 5, a real data smoothing problem is presented by modeling daily S\&P 500 index returns with a stochastic volatility model.

\section{Filtering and Smoothing with SMC}
Consider a general state-space model defined at discrete times $t=1,\ldots,T$:
     \begin{eqnarray*}
     	\text{Initial} :           ~ x_0  & \sim & p(x_0 \given \theta),\\
     	\text{Evolution} :     ~ x_t   & \sim & p(x_t \given x _{t-1},\theta),\\
     	\text{Observation} : ~ y_t  & \sim  &p(y_t \given x_t,\theta), 
     \end{eqnarray*}
   where $y_t$ is the observation, $x_t$ is the hidden state, and $\theta$ are the model parameters.  The Bayesian model is completed with a prior distribution, $\theta \sim p(\theta)$. The state-space model is characterized by two properties: (1) the states $x_t$ follow a first-order Markov process; and (2) the observations are conditionally independent given the states. 
        
    In a Bayesian framework, the objective is to compute the joint posterior distribution of the states and parameters, $p(x_t,\theta|y^s)$, where $y^s=(y_1,\ldots,y_s)$ denotes the observations up to time $s$. When $s=t$, this is called the filtering problem; and when $s=T$, this is called the smoothing problem.  In most models, the joint posterior distribution is unavailable in closed form, and we rely on Monte Carlo methods to sample from the filtering and smoothing distributions.  The goal of this paper is to draw samples from the joint smoothing distribution $p(x^T,\theta|y^T)$.
    
   Traditionally, Sequential Monte Carlo methods assume that $\theta$ is known and are designed to approximate  $p(x_t\given y^t,\theta)$ with a set of weighted samples or particles. In comparison to MCMC methods, SMC avoids convergence problems and allows for efficient calculation of marginal likelihoods, which is useful in parameter estimation or model selection problems.

   The subsections below give a brief review of sampling importance resampling (SIR) particle filters, particle filters with unknown parameters, and the particle learning and smoothing algorithm.   Our new smoothing algorithms are formulated based on this previous work.
  
\subsection{Particle Filtering}

The particle filter was first introduced by \cite*{Gordon:1993} to conduct state estimation in nonlinear/non-Gaussian state-space models. Based on importance sampling, we simply propagate the particles $x_{t-1}^{(i)}$ forward through the system equation and resample the new particles $\tilde{x}_t^{(i)}$ with weights $\omega_t^{(i)}$ proportional to the likelihood $p(y_t\given \tilde{x}_t^{(i)})$ to get filtered particles at time $t$: $x_t^{(i)}$.   The filtering density $p(x_t \given y^t)$ can then be approximated by the empirical density of these particles.
   \begin{equation}
   \begin{split}
   p(x_t\given y^t) 
   & \propto p(y_t\given x_t)\int p(x_t\given x_{t-1})p(x_{t-1}\given y^{t-1})dx_{t-1}\\
   & \approx p(y_t\given x_t) \sum_{i=1}^{N} p(x_t\given x_{t-1}^{(i)}) \omega_t^{(i)}.
   \end{split}
   \end{equation}
   

\subsection{Particle Filtering with Unknown Parameters}

To deal with particle filtering with unknown parameters, \cite*{Kitagawa:1998} introduced the idea of augmenting the state by the parameters as $z_t=(x_t,\theta)'$, then applying a bootstrap filter  to the augmented state vector $z_t$. Moreover, \cite*{Kitagawa:2001} proposed to add noise to the parameters in the transition density to avoid the collapse of samples as time progresses.
    
\cite*{West:2001} proposed an improvement to Kitagawa's method by drawing samples of the parameter from a smoothing kernel density of the form: 
  $$
  p(\theta_{t+1} \given y^t) \approx \sum_{j=1}^{N} w_t^{(j)} \N(\theta_{t+1}\given m_t^{(j)},h^2V_t),
  $$
at each filter step $t$, in which $m_t^{(j)} = a \theta_t^{(j)} + (1 - a)\bar{\theta_t}$, where $\bar{\theta}_t$ and $V_t$ are the sample mean and variance-covariance matrix of the posterior samples of $\theta_t$ at time $t$, and $a=\sqrt{1 - h^2}$ is a smoothing parameter between 0 and 1.  Notice that $a=1$ implies the evolution equation $\theta_{t+1}=\theta_t$, which corresponds to state augmentation with no evolution noise. With this method, we have $V(\theta_{t+1}\given y^t) = V(\theta_{t}\given y^t)$ and thus, no information is lost over time.

  In situations where the posterior distribution of $\theta$ depends on sufficient statistics that are easy to update recursively, the methods from \cite{Storvik:2002} can be applied to draw samples from its filtered distribution. We include sufficient statistics for $\theta$ into the state vector and draw samples of $\theta$ based on sufficient statistics at each time point $t$ in the filtering process. By doing this, the impoverishment problem is mitigated, and the true value of $\theta$ can be learned gradually through a filter process. The approach is based on the decomposition:
\begin{equation}
p(x^t,\theta\given y^t) 
\propto p(x^{t-1}\given y^{t-1}) p(\theta \given s_{t-1}) p(x_t\given x_{t-1},\theta)p(y_t\given x_t,\theta)
\end{equation}  
in which $s_t$ are the sufficient statistics for $\theta$. The details are listed below:

\vspace{.7cm}
\noindent \underline{\textsc{\Large Storvik's SIR Filter}}

\noindent For each time $t=1,\ldots, T$:
\begin{enumerate}
	\item Sample $\theta^{(i)} \sim p(\theta \given s_{t-1}^{(i)})$ (for $i=1,\ldots,N$).
	\item Propagate $x_t^{(i)} \sim p(x_t\given x_{t-1}^{(i)},\theta^{(i)})$ (for $i=1,\ldots,N$).
	\item Compute weights $\omega_t^{(i)} \propto p(y_t\given x_t^{(i)},\theta^{(i)})$ (for $i=1,\ldots,N$).
	\item Update sufficient statistics $s_t^{(i)} = S(x_t^{(i)},s_{t-1}^{(i)},y_t)$ (for $i=1,\ldots,N$).
	\item Resample $N$ times from $\{(x_t^{(i)},s_t^{(i)})\}_{i=1}^N$ with weights  $\omega_t^{(i)}$, to obtain a sample from $p(x_t,s_t\given y^t)$.
\end{enumerate}


\subsection{Particle Learning and Smoothing (PLS)}
In particle smoothing with unknown parameters, we are interested in estimating the states and parameters conditional on the whole data $y^T$ and drawing samples $(x^{T(i)},\theta^{(i)})$ from the joint posterior $p(x^T,\theta \given y^T)$, where $T$ denotes the number of time steps. 

\cite{Carvalho:2010} showed that a backward pass can be run after the filtering and learning algorithm, and the filtered particles could be resampled to obtain draws from the smoothing distribution. The idea is based on Bayes' Rule and the decomposition of the joint posterior smoothing distribution as 
\begin{equation}
p(x^T,\theta \given y^T) = p(x_T,\theta\given y^T) \prod_{t=1}^{T-1} p(x_t\given x_{t+1},\theta,y^t),
\end{equation}
where
\begin{equation}
p(x_t\given x_{t+1},\theta,y^t) \propto p(x_{t+1}\given x_t,\theta) p(x_t\given \theta,y^t).
\end{equation}

The steps of this algorithm are listed below. 

\vspace{.5cm}
\noindent \underline{\textsc{\Large{PLS Algorithm}}}
   \begin{enumerate}
   	\item {\bf (Forward Filter)} Run the particle learning algorithm to generate samples $\{(x_t^{(i)},\theta^{(i)})\}_{i=1}^N$ from $p(x_t,\theta\given y^t)$ at each time $t=1,\ldots,T$.
   	\item {\bf (Backward Smoother)} Select a pair $(x_T^{(i)},\theta^{(i)})$ from Step 1, and simulate backwards: For $t$ = $T-1,\ldots,1$, resample the particles $\{x_t^{(j)}\}_{j=1}^N$ from Step~1 with weights proportional to $\omega_t^{(j)} = p(x_{t+1}^{(i)}\given x_t^{(j)},\theta^{(i)})$ to generate $x_t^{(i)}$.
   \end{enumerate}

According to these authors, this algorithm is an extension of \cite*{Godsill:2004} to state-space models with unknown parameters. However, this is not the case. Note that in the backward pass, we select a fixed $\theta^{(i)}$ first and evaluate the filter weights proportional to $p(x_{t+1}^{(i)}\given x_t^{(j)},\theta^{(i)})$. Thus, correspondingly, we should use samples drawn from $p(x_t\given \theta^{(i)},y^t)$, i.e. the filter samples with respect to this fixed $\theta^{(i)}$. But this is not the case for PLS. 

\begin{figure}
      	\centering
      	\includegraphics[scale = 0.4]{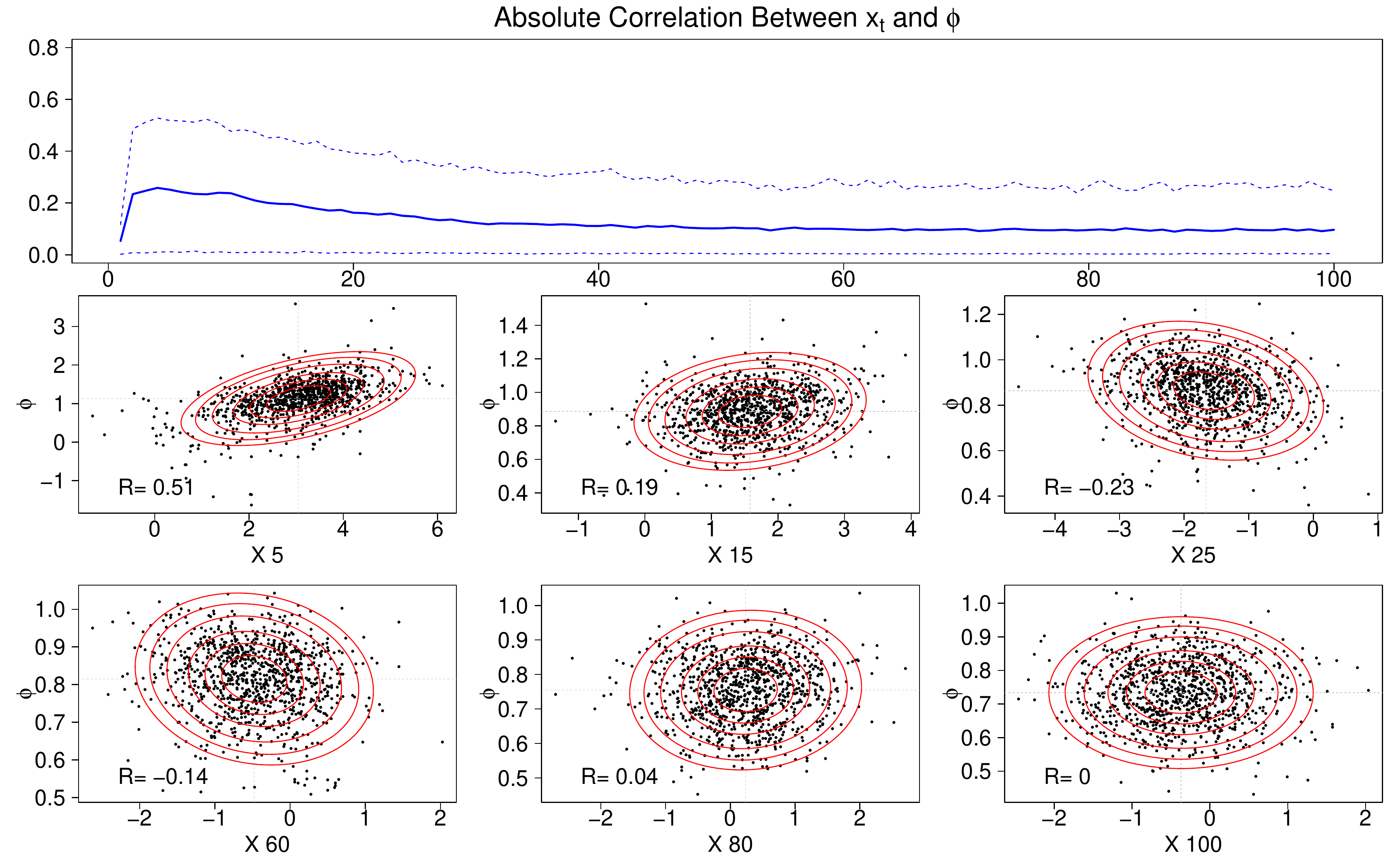} 
      	\caption{\footnotesize AR(1) plus Noise Model. Top: means and 95\% intervals for the posterior absolute correlation between state and AR coefficient $\phi$ at every time step $t$.   Results are based on 500 simulated datasets.  Middle and Bottom: posterior samples of $x_t$ and $\phi$ at selected time steps for one dataset.  The contours represent the fitted normal densities used in the PLS$_a$ algorithm.}
      	\label{fig:AR-corr}
      \end{figure}

Moreover, the particles in the filter process are in fact coming from the marginal density $p(x_t\given y^t)$, not from the conditional density $p(x_t\given \theta,y^t)$. Reweighting particles using the transition density ignores the dependence between states and parameters, which causes inaccurate smoothing estimates when the dependency is strong. Figure~\ref{fig:AR-corr} shows the dependency between the filtered samples of the states and parameters for the AR(1) plus noise model presented in Section 4.  Correlations greater than 0.5 can be detected at the beginning of the time series. The simulation studies presented in Section 4 show that PLS gives poor smoothing estimates, particularly at early periods in the time series.

In the next section, we present two new smoothing algorithms.  The first relies on a transformation of equation (4) and an adjustment of the weights in the backward pass to refine PLS.    The second involves a separate forward-backward pass conditional on the sampled parameters.

\section{Two New Smoothing Algorithms}

\subsection{PLS with Adjustment (PLS$_a$)}
As stated earlier, the PLS algorithm assumes we have samples from the conditional distribution, $p(x_t|\theta,y^t)$ in the filtering algorithm, when in fact we have samples from the joint, $p(x_t,\theta|y^t)$, and hence the marginal, $p(x_t|y^t)$.   Thus, the reweighting scheme in PLS does not give us samples from the target smoothing distribution.  To provide a remedy for this, we consider the following rearrangement of equation (4):
\begin{equation}
\underbrace{p(x_t\given x_{t+1},\theta,y^t)}_{\text{smoother}}
\propto \underbrace{p(x_{t+1}\given x_t,\theta) \; \dfrac{ p( x_t\given \theta,y^t)}{p(x_t\given y^t)}}_{\text{weights}} \underbrace{p(x_t\given y^t)}_{\text{filter}}.
\end{equation}
With samples from the filter, we use $\omega_t^{(j)} = p(x_{t+1}\given x_t^{(j)},\theta) p( x_t^{(j)}\given \theta,y^t)/p(x_t^{(j)}\given y^t)$ as our resampling weights in the backward pass.   Only in this way can we use the filtered particles in the smoothing algorithm.   Note that, in most cases, we cannot compute these resampling weights exactly, since the joint filtering distribution $p(x_t,\theta|y^t)$ is generally not available in closed form.   To fix this problem, we propose to use a multivariate normal approximation to $p(x_t,\theta|y^t)$ based on the filtered particles $\{(x_t^{(i)},\theta^{(i)})\}_{i=1}^N$, using appropriate transformations if necessary.  

The algorithm proceeds exactly as in PLS, but with modified weights in the backward pass.    The details of the particle learning and smoothing algorithm with adjustment (PLS$_a$) are presented below.   Based on the simulation results in Section 4, we find that the adjustment of the weights matters: the adjusted version outperformed the original one significantly, especially in the beginning of the series, where the PLS usually has problems.

\vspace{.5cm}
\noindent \underline{\textsc{\Large{PLS$_a$ Algorithm}}}
\begin{enumerate}
	\item {\bf (Forward Filter)} Run a filtering and learning algorithm to obtain samples $\{(x_t^{(i)},\theta^{(i)})\}_{i=1}^N$ from $p(x_t,\theta \given y^t)$ for $t=1,\ldots,T$.  Use the filtered samples to construct a multivariate normal approximation at each time $t$: 
	$$
	p(x_t,\theta \given y^t) 
	\approx \N\left(\begin{pmatrix} \mu_t^{x} \\ \mu_t^{\theta} \end{pmatrix},
	          \begin{pmatrix} \Sigma_t^{x} & \Sigma_t^{x\theta} \\  \Sigma_t^{\theta x} & \Sigma_t^{\theta} \end{pmatrix}\right).
	$$
	This implies that the marginal and conditional distributions are also normal: 
	$p(x_t\given y^t) \approx \N(\mu_t^{x},\Sigma_t^{x})$, and $p(x_t\given \theta,y^t) \approx \N(\mu_t^{x\given \theta},\Sigma_t^{x \given \theta}),$
	where the conditional mean and covariance are given by the well-known formulas for multivariate normal distributions.
	
	\item {\bf (Backward Smoother)} Select a pair $(x_T^{(i)},\theta^{(i)})$ from Step 1, and simulate backwards: For $t$ = $T-1,\ldots,1$, resample the $\{x_t^{(j)}\}_{j=1}^N$ from Step~1 with weights proportional to 
	$$\omega_t^{(j)} = p(x_{t+1}^{(i)}\given x_t^{(j)},\theta^{(i)})\left( \frac{ p( x_t^{(j)}\given \theta^{(i)},y^t)}{p(x_t^{(j)}\given y^t)} \right)
	                  \approx p(x_{t+1}^{(i)}\given x_t^{(j)},\theta^{(i)})\left( \frac{\N(x_t^{(j)}\given \mu_t^{x \given \theta^{(i)}},\Sigma_t^{x \given \theta} )}{\N(x_t^{(j)}\given \mu_t^{x},\Sigma_t^{x})}\right)$$
	                   to generate $x_t^{(i)}$.   
\end{enumerate}

\subsection{Refiltering Smoothing Algorithm}
  In addition to the the PLS$_a$ modification, we propose the following new smoothing algorithm. The idea is simple but proved to be efficient and accurate in simulation studies. The algorithm is based on the decomposition:
  \begin{equation}
  p(x^T,\theta\given y^T)  =  p(x^T \given y^T,\theta) \; p(\theta \given y^T).
  \end{equation}
We run Storvik's forward filter, or more generally a filter method as in \cite{West:2001}, to get samples of the parameter at the last time step, i.e. $\theta^{(i)} \sim p(\theta\given y^T)$. Then for each $\theta^{(i)}$, we apply a forward-backward smoothing algorithm as in \cite{Godsill:2004} to get one state trajectory $x^{T(i)}$ from $p(x^T\given y^T,\theta^{(i)})$. Repeating this process for each $i$, we obtain states from the marginal smoothing density  $p(x^T\given y^T)$.
 
  Since the run time for the forward filter is negligible compared to the backward smoother, ($O(N)$ vs $O(N^2)$, respectively), in simulation studies, we found that this algorithm almost has the same speed as PLS, but with significant improvement in accuracy. The algorithm is:

\vspace{.5cm}
\noindent  \underline{\textsc{\Large Refiltering Algorithm}}	
  \begin{enumerate}
  	\item {\bf (Forward Filter)} Use Storvik, Particle Learning or Liu \& West to run a forward filter and learning algorithm and generate $\theta^{(i)} \sim p(\theta\given y^T)$;
  	\item {\bf (Backward Smoother)} For each $\theta^{(i)}, i=1,\ldots,N_0$, run a forward-backward smoothing algorithm to get a sample $x^{T(i)} \sim p(x^T\given y^T,\theta^{(i)})$. 
  \end{enumerate}         

  Note that this algorithm has a complexity of $O(TN^2)$, the same as PLS. But we can make it an $O(TN)$ algorithm in two ways. The first one is that we can choose a small number of states $n_0 \ll N$ for the forward-backward smoother in step 2. The second is to use a small number of parameter draws of size $N_0$ to use in step 2. The simulation study showed that both methods make the algorithm run much faster with only a minor loss of accuracy. 

  In the case where the model is linear and Gaussian: $x_t = G_t x_{t-1} + w_t,  \enspace w_t \sim \N(0,W)$;
    $y_t  = F'_t x_t + v_t, \enspace v_t \sim \N(0,V)$; 
  we can incorporate a forward filtering, backward sampling algorithm as in \cite{Carter:1994} and \cite{Fruhwirth:1994} into step 2 : we run a Kalman filter forward pass then generate a sample backwards based on equation $(3)$. Note $p(x_t \given y^{t-1}) \sim \N(a_t,R_t)$ is the prior and $p(x_t \given y^t) \sim \N(m_t,C_t)$ is the posterior of the state at each time point $t$, which depends on the parameters $\theta=(F_t,G_t,V,W)$.

\vspace{.5cm}
\noindent  \underline{\textsc{\Large Refiltering with FFBS}}
  \begin{enumerate}
  	\item {\bf (Filter)} Use Storvik, Particle Learning or Liu \& West to run a forward filter and learning algorithm and generate $\theta^{(i)} \sim p(\theta\given y^T), i=1,..,N$;
  	\item  {\bf (Smoother)} For each $\theta^{(i)}, i=1,..,N_0$, run a Kalman filter and store prior and posterior moments $a_t,R_t,m_t,C_t$. Sample  $x_T^{(i)} \sim \N(m_T,C_T)$. For $t = T-1$ to 1, sample  $x_t^{(i)} \sim p(x_t\given x_{t+1}^{(i)},\theta^{(i)},y^t) = \N(h_t,H_T)$, in which $h_t = m_t + B_t(x_{t+1}^{(i)} - a_{t+1})$, $H_t = C_t - B_t R_{t+1} B'_t$ and $B_t = C_t G'_{t+1} R^{-1}_{t+1}$. This provides a sample, $x^{T(i)} \sim p(x^T\given y^T,\theta^{(i)})$. 
  \end{enumerate}

  \section{Examples}
   
  \subsection{AR(1) Model with Three Unknown Parameters}
  Assume that the states $x_t$ follow an AR(1) process where the observations $y_t$ equal $x_t$ plus Gaussian noise:
  $$x_t = \phi  x_{t-1} + w_t, \enspace w_t \sim \N(0,W),$$
  $$y_t = x_t + v_t, \enspace v_t \sim \N(0,V).$$
  This benchmark model has been widely used in SMC and MCMC simulation studies \cite*[see, for example,][]{Storvik:2002,Polson:2008}. 
  In this model, FFBS can be easily implemented and a long chain MCMC with 150,000 iterations was set as a standard to compare with other smoothing algorithms.  We generate $T=100$ observations with parameter values $V = W = 1$, $\phi = 0.75$ and $x_0=0$.

  For the analysis, we assume conjugate priors for the parameters: $(\phi,W) \sim \NIG(b_0,B_0,n_0,d_0)$, and $V \sim  \IG(\nu_0,\delta_0)$ where $\IG(a,b)$ denotes the inverse-gamma distribution with scale and shape parameters $a$ and $b$, and $\NIG$ denotes the normal-inverse gamma distribution where $B_0$ represents the inverse of the scale factor in the normal variance.   We assume $n_0 = \nu_0 = d_0 = \delta_0 = 2$ and $b_0=0.5, B_0=1$.  The conjugate model for the parameters allows us to use Storvik's algorithm, with the sufficient statistics $s_t=(b_t,B_t,n_t,d_t,\nu_t,\delta_t)$, and the updating recursions:
   \begin{align*}
  B_t &=  B_{t-1} + x_{t-1}^2, &
  b_t &= B_t^{-1}(B_{t-1}b_{t-1}+ x_{t-1}x_t),\\
  n_t &= n_{t-1} + 1/2, &
  d_t &= d_{t-1} + (b_{t-1}^2B_{t-1} + x_t^2 - b_t^2B_t)/2,\\
  \nu_t &= \nu_{t-1} + 1/2, &
  \delta_t &= \delta_{t-1} + (y_t - x_t)^2/2.
  \end{align*} 
    
\begin{figure}
      	\centering
      	\includegraphics[scale = 0.4]{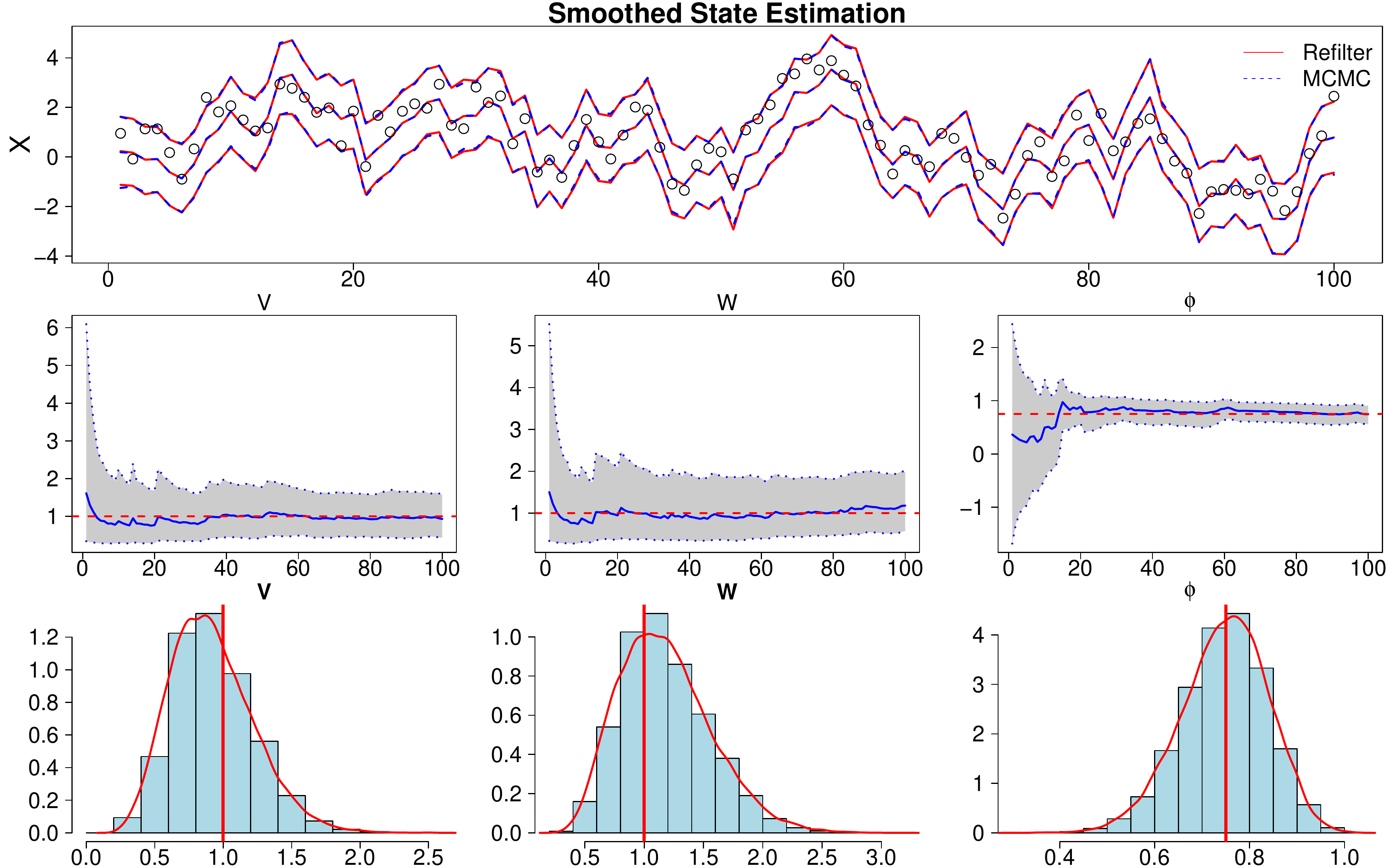} 
      	\caption{\footnotesize AR(1) Plus Noise Model. Top: posterior mean and the $2.5^{th}$, $97.5^{th}$ quantiles based on the refiltering smoother (blue) and full MCMC (red). Middle: parameter learning in Storvik forward filter. Bottom: histograms of parameter samples at last filter step, superimposed  with density estimation from long MCMC (red line).   The true parameter values are indicated by horizontal and vertical red lines.}
      	\label{fig:AR-learn}
      \end{figure}

  We first run Storvik's filtering algorithm. Figure \ref{fig:AR-learn} shows the parameter learning plots and the posterior distribution at the last time period $T = 100$. From the plot, we notice the true parameters values were learned properly and the samples of the parameters at the last time step are well concentrated around the true parameter values. Also the samples from the filter agree well with samples from a long MCMC. State smoothing by refiltering and the result of a long MCMC are also presented in Figure \ref{fig:AR-learn}. We notice that the mean, $2.5^{th}$ and $97.5^{th}$ quantiles of the smoothing samples almost coincide for the two methods at each time step $t$.
   
\begin{figure}
      	\centering
      	\includegraphics[scale = 0.35]{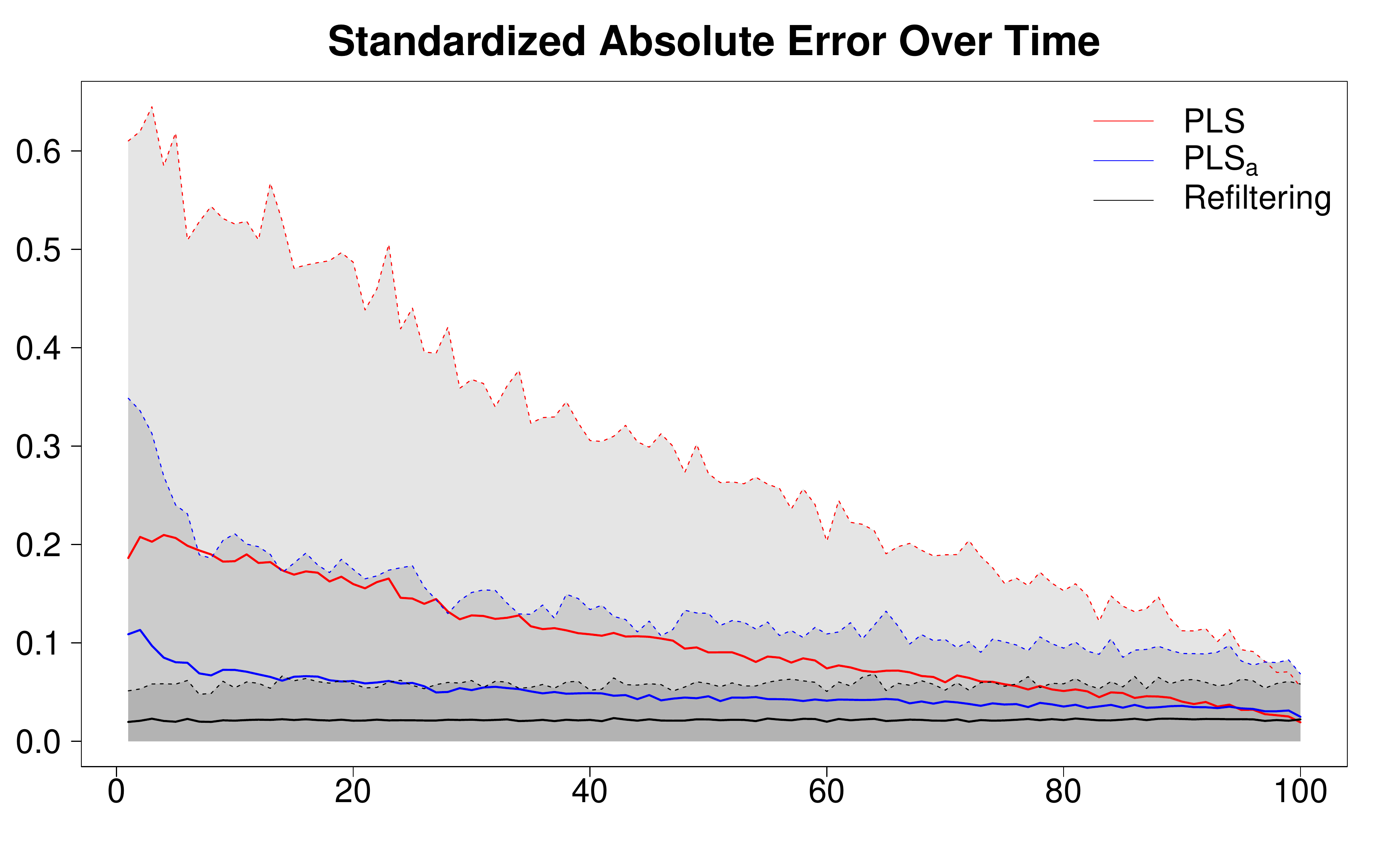} 
      	\caption{\footnotesize AR(1) Plus Noise Model. Standardized absolute errors over time for three smoothing algorithms compared to a long MCMC. The results are based on 500 simulations. The long MCMC is treated as the truth. The solid lines are the means of standardized absolute errors at time $t$ among 500 simulations, while the dashed lines represent the $95^{th}$ quantile.}
  	   	\label{fig:AR-MAE}
      \end{figure}

   To show that PLS$_a$ outperforms PLS using the same computation time, we ran 500 simulations for each of these two methods and compared the standardized absolute errors over time, i.e., $\hat{e}^*_t = |\hat{x_t} - \hat{x}^{true}_t|/\sigma(x_t\given y^T)$ for $t = 1 \dots T$, where $\hat{x}_t^{true}$ and $\sigma(x_t|y^T)$ are the smoothed mean and standard deviation for $x_t$ computed from the long MCMC, and $\hat{x}_t$ is the smoothed mean from the other algorithms. The result is shown in Figure \ref{fig:AR-MAE}. From the plot, we can see the main difference between the two smoothing algorithms appears at the beginning of the series, in which the dependence of states and parameters is strong and therefore the adjustment matters. As time progresses, the dependency of states and parameters decreases, and the adjusted smoothing outcomes coincide with PLS. The results from the refiltering smoothing algorithm are also shown in the plot. For this model, refiltering substantially outperforms the other two methods, and its accuracy is consistent over time.  Note that the number of particles for the three smoothing algorithms was adjusted to assure similar computation time.

   To compare the performance of all of the smoothing algorithms in this paper, we implement long and short MCMC runs, PLS, PLS$_a$, refiltering, $O(TN)$ refiltering, and refiltering with FFBS using 500 data simulations.   All SMC based smoothing algorithms are run in parallel on 16 cores on a single node. Based on a similar run time, the mean standardized absolute errors over time (MAE* = $\sum_{t=1}^T|\hat{e}^*_t|/T$) are listed in Table~\ref{table:AR-MAE}. From the table, we see that the MAE* values for PLS$_a$ are about half as large as for PLS. For all of the refiltering algorithms, the MAE* magnitude is only about one fifth of of that for PLS.  Hence, both of the new smoothing algorithms outperform the PLS smoothing algorithm of \cite{Carvalho:2010}.  The column labeled MAEP* represents the mean standardized absolute error between the posterior mean of the parameters at the last time step for a long MCMC versus the other algorithms.  From the table, we see that the learning of parameters using the particle filter is almost as good as the learning from a short MCMC.

	\begin{table}
		\centering
				\caption{Comparison of smoothing algorithms in AR(1) plus noise model.   MAE* and MAEP* denote the standardized mean absolute error for the states
				and parameters, respectively.  SMC smoothers are based on Storvik's algorithm with $N=50,000$ particles.} 
		\begin{tabular}{lcccc} 
			\hline\hline
			\rule{0pt}{2.5ex}  Algorithm            &  $N (N_0/n_0)$   & Time         & MAE*  & MAEP* \\ \hline 
			\rule{0pt}{2.5ex} MCMC                 & 5000           & 17s             & 0.019  & 0.051    \\ 
			\rule{0pt}{2.5ex} PLS                      & 2300          & 22s             & 0.138   & 0.058  \\ 
			\rule{0pt}{2.5ex} PLS$_a$              & 1050          & 22s             & 0.060   & 0.058  \\ 
		      \rule{0pt}{2.5ex} Refiltering            & 1,500/1500  & 22s             & 0.026   & 0.058  \\ 
	              \rule{0pt}{2.5ex} Refiltering            & 10,000/150  & 22s             & 0.022   & 0.058  \\ 
		      \rule{0pt}{2.5ex} Refiltering            & 1,000/2500  & 23s             & 0.031   & 0.058  \\ 
			\rule{0pt}{2.5ex} Refiltering/FFBS  & 44,000         & 21s             & 0.015   & 0.058  \\ 
			\hline\hline   &  \\
		\end{tabular}
		\label{table:AR-MAE}
	\end{table}

   \subsection{Nonstationary Growth Model with Five Unknown Parameters}
   Consider the nonstationary growth model:
   \begin{align*}
   x_t &= \alpha x_{t-1} + \beta \frac{x_{t-1}}{1+x_{t-1}^2} + \gamma \cos(1.2(t-1)) + w_t,\\
   y_t & = x_t^2/20 + v_t,
   \end{align*}
in which $w_t \sim \N(0,W)$ and $ v_t \sim \N(0,V)$.   This benchmark nonlinear time series model has been used by \cite{Carlin:1992} to test MCMC smoothing, by \cite{Gordon:1993} to test the bootstrap filter, and by \cite{Doucet:2010} to test the Forward-Backward smoothing with known parameters. Here we test our smoothing methods on this model with unknown parameters.

\begin{figure}
           	\centering
           	\includegraphics[scale = 0.4]{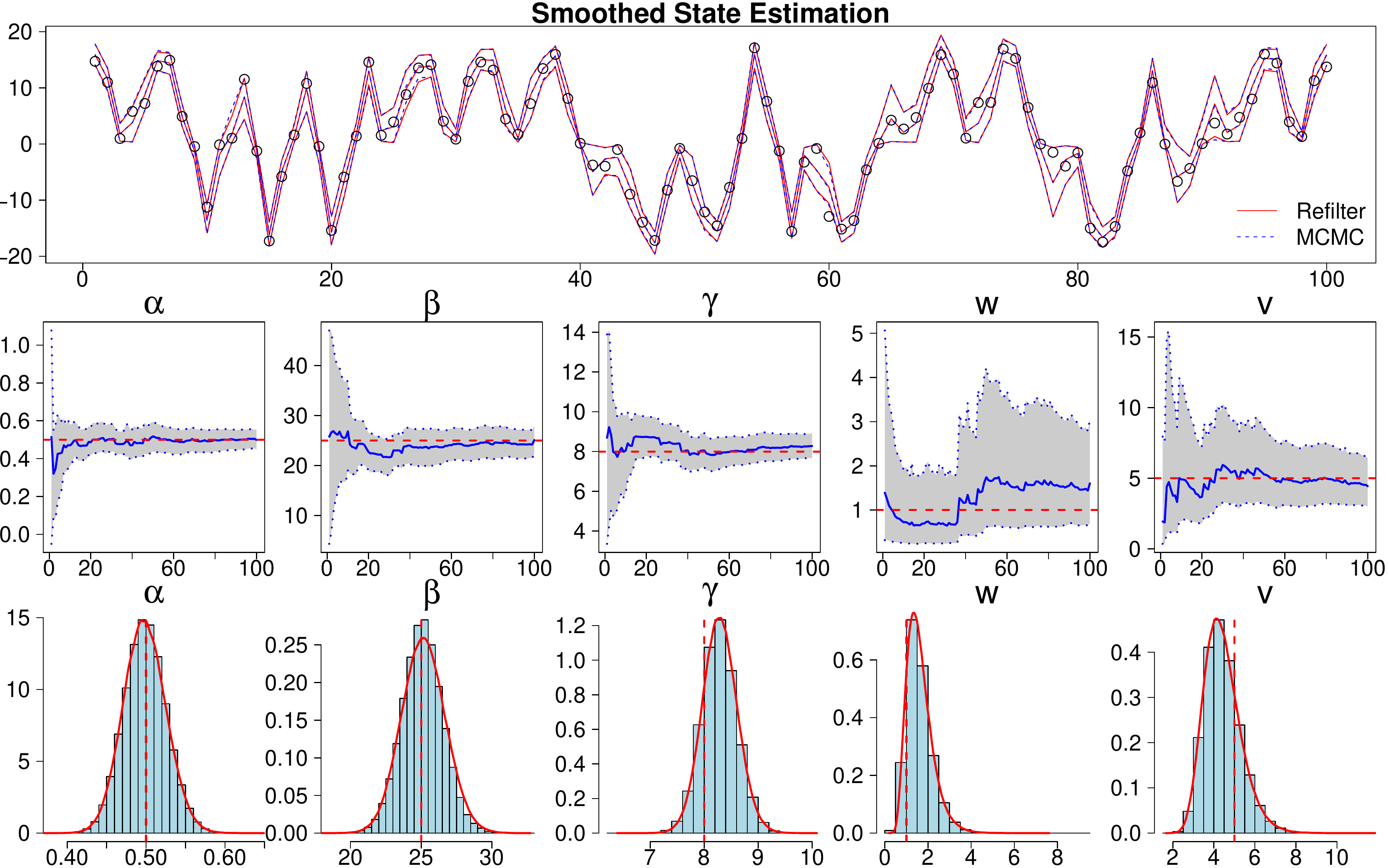} 
            \caption{\footnotesize Nonlinear Model.
            	Top: refiltering smoothing compared with MCMC.
            	Middle: parameter learning in Storvik forward filter. 
            	Bottom: histogram of parameter samples at last filter step, superimposed  with density estimation from long MCMC (red line).}
           	\label{fig:NL-learn}
           \end{figure}

   We generate $T=100$ observations using parameter values $\alpha = 0.5$, $\beta = 25$ , $\gamma = 8$, $V =5$ and $W = 1$.  We assume conjugate priors for the parameters similar to those given in \cite{Carlin:1992}, i.e. $((\alpha,\beta,\gamma)',W) \sim \NIG(b_0,B_0,n_0,d_0)$, and $V \sim  \IG(\nu_0,\delta_0)$, where $b_0 = (0.5,25,8)'$, $B_0 = \text{diag}(1/0.25^2,1/10^2,1/4^2)$, and $n_0 = \nu_0 = d_0 = \delta_0 = 2$.  The conjugate priors allow us to use Storvik's algorithm for filtering and parameter learning.  The updating recursions for the sufficient statistics are given by 
   \begin{align*}
  B_t &= B_{t-1} + F_t F_t', &
  b_t &= B_t^{-1} (B_{t-1}b_{t-1} + F_t x_t),\\
  n_t &= n_{t-1} + 1/2, &
  d_t &= d_{t-1} + (b_{t-1}'B_{t-1}b_{t-1} + x_t^2 - b_t'B_tb_t)/2,\\
  \nu_t &= \nu_{t-1} + 1/2, &
  \delta_t &= \delta_{t-1} + (y_t - x_t^2/20)^2/2,
  \end{align*} 
where $F_t=(x_{t-1},x_{t-1}/(1+x_{t-1}^2),\cos(1.2(t-1))$. The parameter learning process and the posterior histograms of the parameters at time $T=100$ are plotted in Figure \ref{fig:NL-learn}. In the figure, the 95\% confidence bands narrow down quickly as time increases. In the histograms, the samples concentrate around the true parameter values. A total of $N = 50,000$ particles were used for the forward pass.
  
 Furthermore, we compare the refiltering smoothing algorithm using $N_0=5000$ and $n_0 = 1000$, with a long MCMC using $N = 150,000$ iterations. The smoothing plot is also presented in Figure \ref{fig:NL-learn}. The results from the two smoothing algorithms closely agree with each other. 
  
   Table~\ref{table:NL-MAE} gives a summary of the overall performance of the three smoothing algorithms compared to a long MCMC, using 500 simulated datasets. A decrease in the mean absolute error for the new methods relative to PLS is obvious. The plot of the standardized absolute errors over time of the three smoothing algorithms (not shown) illustrates the same patterns as for the AR(1) model: the main improvement of the two new smoothing algorithms over PLS is evident at the beginning of the time series.
   
   Note that for this model, it is difficult to distinguish between the positive and negative sign of the states based on the data, thus it is difficult to assign initial values for the states for the MCMC algorithm based on observations. With a bad starting values for the states, the MCMC chain takes much longer to converge. In contrast, smoothing based on SMC does not suffer from the initialization problem.

   	\begin{table}
   		\centering
   		\caption{Comparison of smoothing algorithms for the nonlinear growth model.   
		SMC smoothers are based on Storvik's algorithm with $N=50,000$ (run time 96s).}
   		\begin{tabular}{lcccc}
   			\hline\hline
   			\rule{0pt}{2.5ex}  Algorithm          & $N (N_0/n_0)$                 &  Time          & MAE*  & MAEP* \\ \hline 
   			\rule{0pt}{2.5ex} MCMC               & 20000               & 228s          & 0.075  & 0.246   \\ 
   			\rule{0pt}{2.5ex} PLS                   & 10000                & 200s    & 0.373  & 0.213   \\ 
   			\rule{0pt}{2.5ex} PLS$_a$           & 5000                  & 208s    & 0.189  & 0.213   \\ 
   			\rule{0pt}{2.5ex} Refiltering         &  5000/1000         & 231s    & 0.097  & 0.213   \\ 
			\hline\hline 
   		\end{tabular}
		\label{table:NL-MAE}
   	\end{table}
     
   \subsection{Chaotic Model with Three Unknown Parameters}
   Now let us consider data generated from the model:
   \begin{align*}
   N_{t} &= r N_{t-1} e^{-N_{t-1} + z_t},  \enspace  z_t \sim \N(0,\sigma^2),\\
   y_t   &\sim Pois(\phi N_t).
   \end{align*}
   This model is widely used in the field of ecology \citep*{Fasiolo:2016}, where $N_t$ stands for the density of the population at generation $t$, and $r$ is the growth rate of the population. This model is characterized by its sensitivity to parameter variations: small increments in $r$ will lead to significant oscillations in the likelihood function. As a result, parameter estimation via maximum likelihood methods is challenging. \cite*{Fasiolo:2016} described the pathological likelihood function for this model and compared the performance of information reduction approaches and state-space methods for this model. A time series of 100 observations is generated from this model with true parameter values $r = e^{3.8}$, $\sigma^2=0.3$ and $\phi = 10$.
   
   To estimate this model using our framework, we first make the transformations, $x_t = \log(N_t)$ and $\mu=\log(r)$.  Then the system and observation equations become
   \begin{align*}
   x_t &= \mu + x_{t-1} - e^{x_{t-1}} + z_t,\\
   y_t & \sim  Pois(\phi e^{x_t}).
   \end{align*}
 
   We assume diffuse conjugate priors for the parameters of the form, $\phi \sim \Ga(a_0,b_0);$ and $(\mu, \sigma^2) \sim \NIG(m_0,c_0,n_0,d_0)$, where $\Ga$ denotes the gamma distribution, with $a_0=15, b_0=1, m_0=5, c_0=.1, n_0=2, d_0=2$. The sufficient statistics are $s_t=(a_t,b_t,m_t,c_t,n_t,d_t)$, and the updating recursions are
   		\begin{align*}
   		a_t & = a_{t-1} + y_t,  &
   		b_t & = b_{t-1} + e^{x_t},	\\
   		c_t & = c_{t-1} + 1, &
		m_t & = c_t^{-1} (c_{t-1}m_{t-1} + x_t - x_{t-1} + e^{x_{t-1}}),\\
   		n_t & = n_{t-1} + 1/2,&
   		d_t & = d_{t-1} + \{c_{t-1}m_{t-1}^2 + (x_t - x_{t-1} + e^{x_{t-1}})^2  - c_tm_t^2 \}/2.
   		\end{align*} 
   		
\begin{figure}
                	\centering
                	\includegraphics[scale = 0.4]{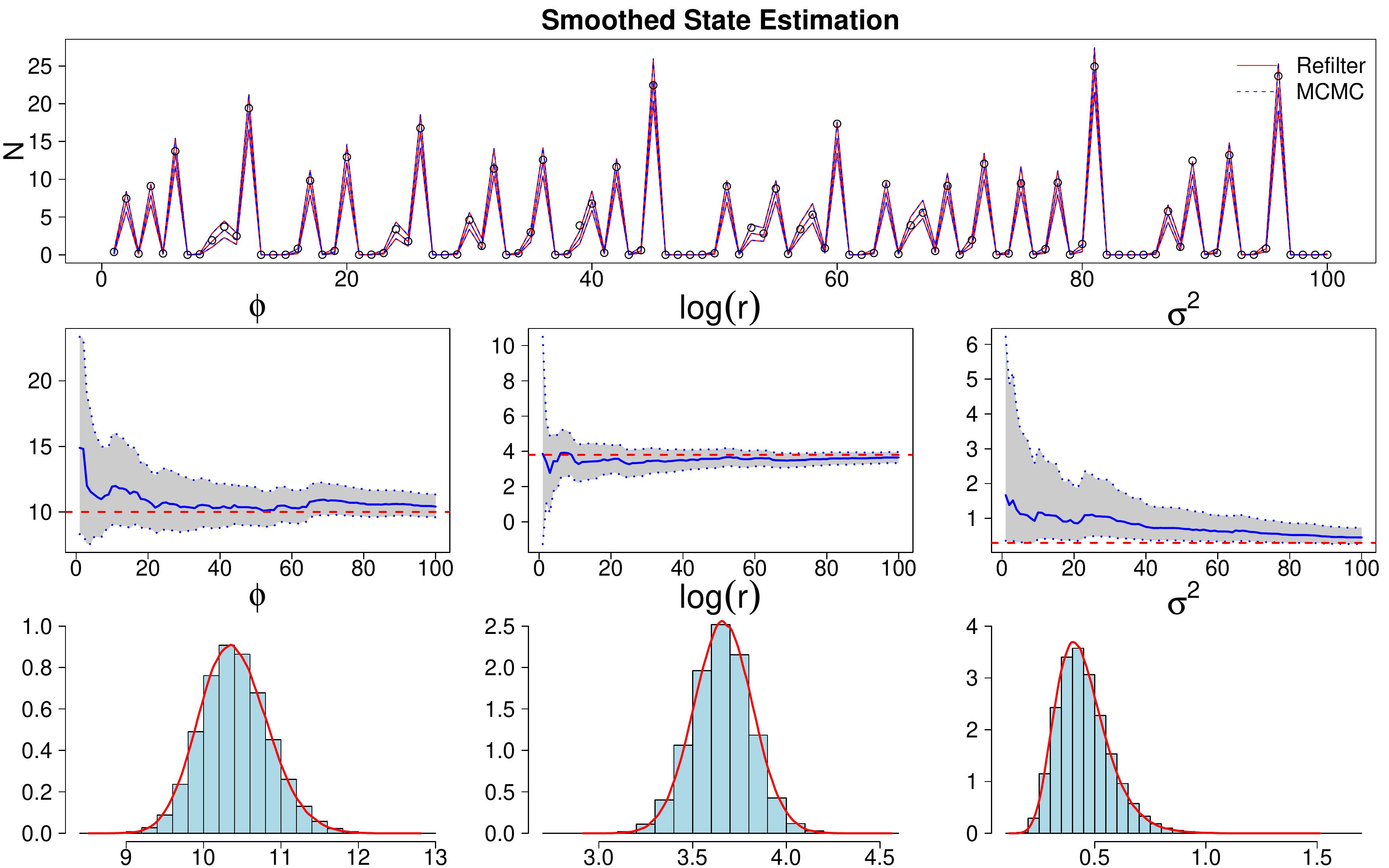} 
                	\caption{\footnotesize Chaotic Model.
                		Top: refilter smoothing compared with MCMC.
                		Middle: parameter learning in Storvik forward filter process. 
                		Bottom: histogram of parameter samples at last filter step, superimposed  with density estimation from long MCMC (red line). }
                	\label{fig:Chaos-learn}
                \end{figure}

   The parameter learning process is summarized in Figure \ref{fig:Chaos-learn}. A total of $N = 50,000$ particles were used for the filtering. The true parameter values were learned quickly and the posterior samples of the parameters settle around the true values. Figure \ref{fig:Chaos-learn} also provides a comparison of refiltering with $N_0=5000$ and $n_0=1000$ to a long MCMC with 150,000 iterations for smoothing, which shows similar results for both methods.
   
   To allow a comparison of the three smoothing methods, 100 simulations were performed. We examined plots of the MAE* values over time (not shown), and Table~\ref{table:Chaos-MAE} presents numerical summaries based on the simulations. From the plots, we find similar patterns for this example as in the previous two: PLS$_a$ and refiltering dominate PLS early in the time series (up to about time $t=80$), and the three methods coincide afterwards. From Table~\ref{table:Chaos-MAE}, we notice a decrease in MAE* for the two new methods compared to PLS. We also find that smoothing method based on refiltering performs better compared to a short MCMC for this model.
                
       \begin{table}
      		\centering
      		\caption{Comparison of smoothing algorithms for the chaotic model.   
			SMC smoothers are based on Storvik's algorithm with $N=50,000$ (run time 11s).}
      		\begin{tabular}{lcccc} 
      			\hline\hline
      			\rule{0pt}{2.5ex}  Algorithm        & $N (N_0/n_0)$                      & Time          & MAE*  & MAEP* \\ \hline 
      			\rule{0pt}{2.5ex} MCMC             & 30000                   & 184s          & 0.097 & 0.166    \\ 
      			\rule{0pt}{2.5ex} PLS                 & 10000                    & 219s    & 0.190 & 0.206    \\ 
			\rule{0pt}{2.5ex} PLS$_a$         & 5000                     & 180s    & 0.108 & 0.206    \\ 
      			\rule{0pt}{2.5ex} Refiltering        & 5000/1000            & 40s      & 0.089 & 0.206   \\ 
			\hline\hline 
      		\end{tabular}
       		\label{table:Chaos-MAE}
      	\end{table}

    \subsection{Analysis of S\&P 500 Returns}
   In this section, we analyze daily returns on the S\&P 500 index from January 2008 to March 2009, during the financial crisis, and compare the PLS, PLS$_a$ and refiltering smoothers with MCMC where  daily returns $y_t$ follow a stochastic volatility model:
  \begin{align*}
   x_{t} &= \alpha + \beta x_{t-1} + \omega_t,  \enspace  \omega_t \sim \N(0,W),\\
   y_t   &= \mu + \exp(x_t/2)v_t, \enspace v_t \sim \N(0,1). 
   \end{align*}
Here $y_t=\log(P_t/P_{t-1})$ are the daily returns, $P_t$ are the prices, $\mu$ is the expected return, and $x_t$ is the unobserved log-variance at time $t$, which is assumed to follow an AR(1) model with drift $\alpha$. The AR coefficient $\beta$ measures the autocorrelation present in the logged squared data.     This model has been widely used to analyze financial time series with volatility clustering \cite*[see, for example,][]{Jacquier:1994,Kim:1998}.
    
We assume conjugate priors for the parameters $\theta=(\mu,\alpha,\beta,W)$.   For the expected returns, $\mu \sim \N(a_0,b_0)$, and for the volatility parameters, we assume $((\alpha,\beta)',W) \sim \NIG(m_0,C_0,n_0,d_0)$, where $a_0=0, b_0=1, m_0=(0,.9)', C_0=\mbox{diag}(1,1), n_0=2, d_0=2$.  The refiltering algorithm is implemented with $N=10,000$ and $n_0 = 1000$. The parameter learning and state smoothing estimates are compared to an MCMC with 15,000 iterations, using the single-state updating scheme of \cite{Jacquier:1994}.

From the sequential learning plots in Figure 6, a significant change in the parameters is observed in September 2008, especially for $\alpha$ and $\beta$.  The change corresponds to the collapse of Lehman Brothers, and an increase in the volatility of the index.  Figure 7 shows the filtered and smoothed volatilities for each algorithm.  These plots show clear evidence that PLS and MCMC do not match, especially from September-November 2008, when the volatility changes abruptly.  PLS$_a$ reduces this discrepancy somewhat, and among the three new smoothing algorithms, refiltering is by far the most accurate.  Given that the run times for refiltering and MCMC are roughly the same, and the close match between the corresponding posterior distributions, we conclude that these two algorithms are comparable.

                \begin{figure}
                	\centering
             	\includegraphics[scale = 0.8]{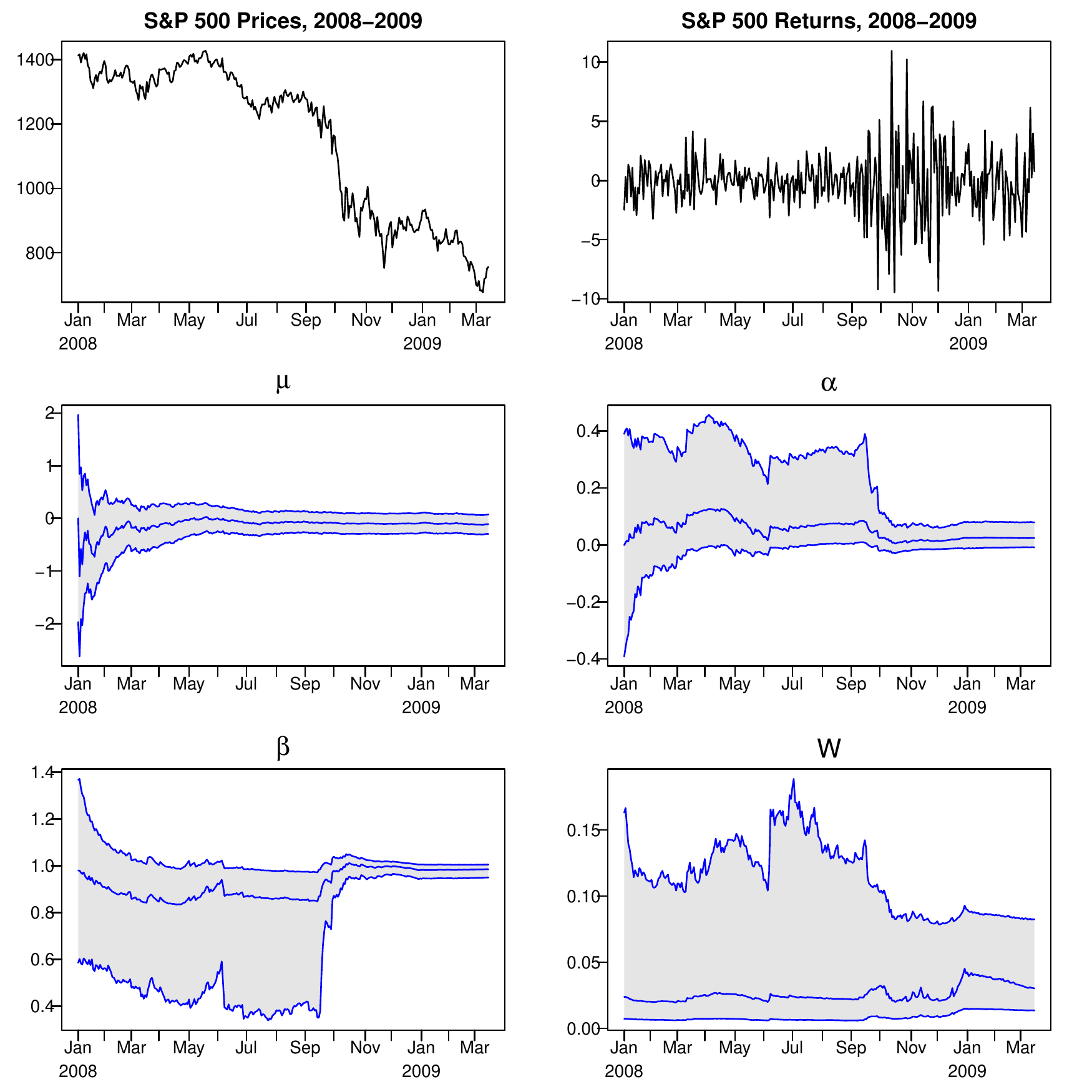}
          	\caption{\footnotesize Stochastic Volatility Model.
       		Top row: Daily prices and returns on the S\&P 500 index from January 2008 to March 2009.
		Middle and bottom rows: filtered medians and 95\% intervals for the parameters $\mu$, $\alpha$, 
		$\beta$ and $W$.  Based on Storvik's algorithm with 50,000 particles.}
               	\label{fig:sv-params}
             \end{figure}

                \begin{figure}
                	\centering
             	\includegraphics[scale = 0.65]{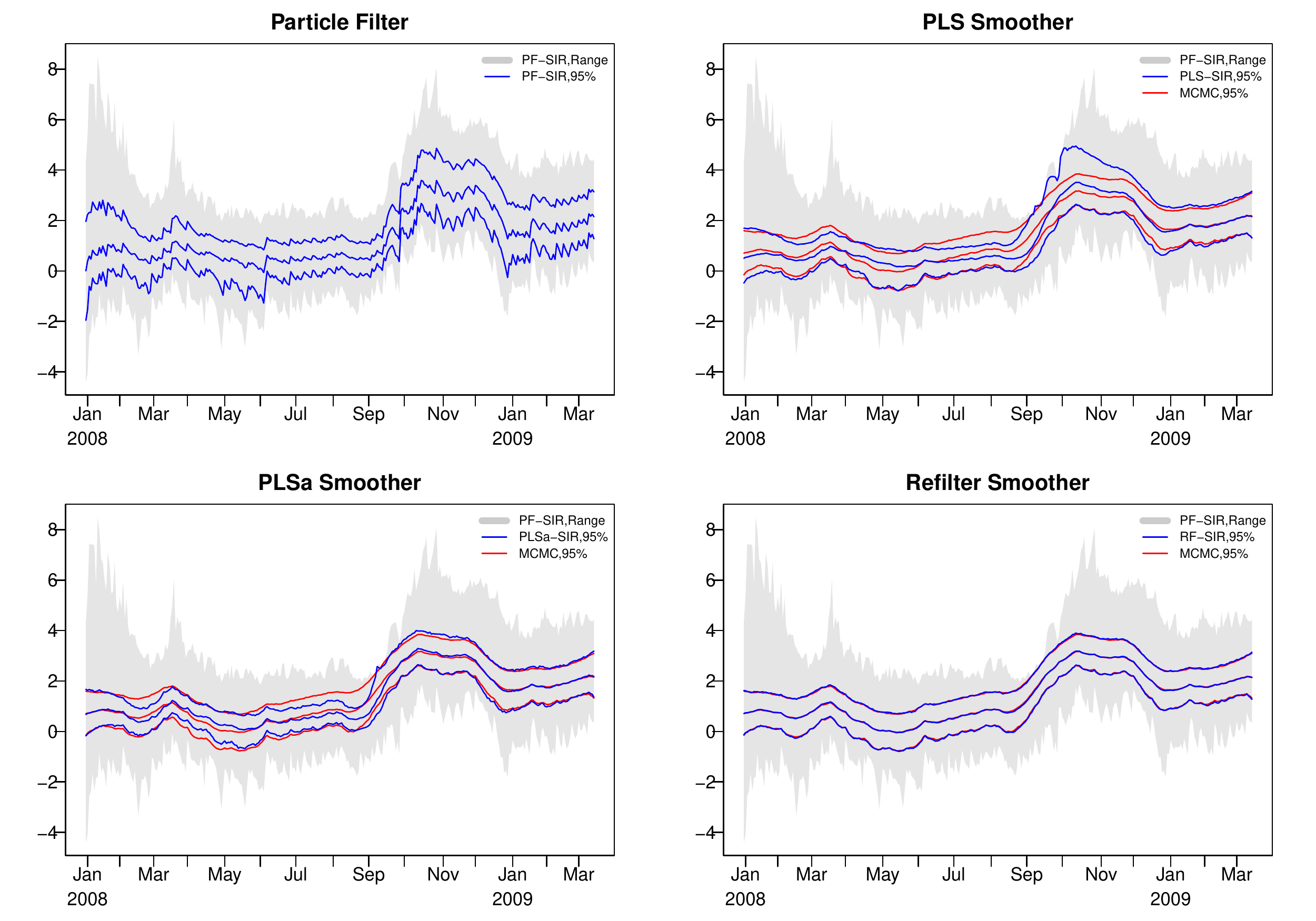} 
          	\caption{\footnotesize Stochastic Volatility Model.
       		Top left: filtered median and 95\% posterior intervals for $x_t$ from Storvik's algorithm. 
		 The other three panels show the smoothed median and 95\% intervals for $x_t$ 
		 using MCMC and the PLS, PLS$_a$ and refiltering algorithms.  The gray bands in
		 each plot are the 100\% filtering intervals (i.e., range) from Storvik's algorithm.}
                	\label{fig:sv-states}
             \end{figure}

   \section{Conclusions}
   
   In this paper, we proposed two new SMC-based smoothing algorithms that simultaneously deal with parameter learning.  The first is a modification of the PLS algorithm of \cite{Carvalho:2010}, that adds a correction term in the backward resampling weights. The second is a two-step algorithm, called {\em refiltering}, that includes a parameter learning step followed by a forward-backward algorithm for smoothing.  Refiltering is well suited for parallel implementation, since the smoothing step requires essentially no communication between processors.  
   We tested the new methods on four models: a benchmark AR(1) plus noise model, a nonlinear growth model, a chaotic model from ecology, and a stochastic volatility model from finance, and compared the estimates with the widely-used smoothing method known as PLS.  For all examples, both new methods showed significant improvement over PLS, and refiltering was competitive with MCMC.  Overall, our proposed methods are quite general, and may be applied to a wide class of state-space models for parameter and state estimation.    In future work, we plan to apply the methods to other real data applications in finance and ecology.
   
   \newpage
    \section*{Appendix A: Marginal Likelihood of the Model}
    The marginal likelihood is important in Bayesian model selection.    As noted by \cite{Carvalho:2010}, the marginal likelihood can be computed trivially from the output of SMC-based Bayesian filtering and learning algorithms \cite[e.g.,][]{Storvik:2002,Carvalho:2010}.  
Define $\omega_t^{j} = p(y_t|x_t^{j},\theta^{j},\M)$, where $(x_t^j,\theta^j) \sim p(x_t,\theta|y^{t-1},\M)$ for given model $\M$.  Then, the log marginal likelihood for model $\M$ is estimated by
       \begin{equation*}
       \log(f(y \given \M)) \approx \sum_{t=1}^{T} \log\left(\sum_{j=1}^{N}\omega_t^{j}\right) - T \log(N).
       \end{equation*}
   By comparison, many different MCMC-based estimates of the marginal likelihood have been proposed.  One of the most commonly used and straightforward methods is the harmonic mean estimator \citep{Newton:1994}, which can be computed based on the joint distribution of the data:
   \begin{equation*}
   \log(f(y \given \M)) \approx \log\left(\dfrac{1}{\frac{1}{N}\sum_{j=1}^{N}\frac{1}{p(y\given \psi^j,\M)}}\right),
   \end{equation*}
   where $y=(y_1,\ldots,y_T)$ are the observations, $N$ is the total number of MCMC iterations, and $\psi^j=(x^j,\theta^j), j=1,\ldots,N$ are the posterior draws of the states and parameters.
   
   An important advantage of SMC over MCMC is that the estimation of marginal likelihood from SMC output is stable.  As shown in Table 4, in our simulation studies, we find that the SMC-based marginal likelihood estimator converges quickly as $N$ increases, while for MCMC, the harmonic mean estimator fails to converge even for $N$ larger than 500K in all three models.

\begin{table}[h]
	\centering
	\caption{Comparison of Log Likelihood}
\begin{tabular}{l l l l l l l l l l}
	\hline
	\rule{0pt}{2.5ex}     &  & \multicolumn{2}{l}{  AR(1) + Noise } &  & \multicolumn{2}{l}{  Nonlinear Growth } &  & \multicolumn{2}{l}{  Chaotic Model } \\\cline{3-4}\cline{6-7}\cline{9-10}
	\rule{0pt}{2.5ex}$N$    &  & Storvik & MCMC                       &  & Storvik & MCMC                          &  & Storvik & MCMC                       \\ \hline\hline
	\rule{0pt}{2.5ex}1K   &  & -176.44 & -164.40                    &  & -212.91 & -253.27                       &  & -286.34 & -193.37                    \\
	\rule{0pt}{2.5ex}5K   &  & -176.71 & -171.74                    &  & -212.51 & -169.92                       &  & -284.88 & -200.35                    \\
	\rule{0pt}{2.5ex}10K  &  & -176.71 & -169.88                    &  & -211.66 & -170.42                       &  & -284.96 & -195.64                    \\
	\rule{0pt}{2.5ex}50K  &  & -176.87 & -169.86                    &  & -212.45 & -170.91                       &  & -285.16 & -199.54                    \\
	\rule{0pt}{2.5ex}100K &  & -176.91 & -170.84                    &  & -212.10 & -162.56                       &  & -284.98 & -196.61                    \\
	\rule{0pt}{2.5ex}500K &  & -176.88 & -173.56                    &  & -212.30 & -164.06                       &  & -285.10 & -197.99                    \\ \hline
\end{tabular}
\end{table}
\label{table:loglike}

	\clearpage
	\bibliographystyle{rss}
	\bibliography{sample}
\end{document}